\documentclass[sigconf,screen]{acmart}

\usepackage{tabularx}
\usepackage{booktabs}
\usepackage{hyperref}
\usepackage{multirow}
\usepackage{listings}

\usepackage{makecell}
\usepackage{graphicx}
\usepackage{subcaption}
\usepackage{wrapfig}
\usepackage{framed}
\usepackage{amsthm}
\usepackage{xcolor}
\usepackage{flushend}
\usepackage{pifont} 
\usepackage{amsmath}
\usepackage{paralist}
\usepackage{siunitx}
\usepackage{enumitem}
\usepackage{algorithm}
\usepackage[noend]{algorithmic}
\usepackage{fancyhdr}

\newcommand{\Space}[1]{}

\colorlet{shadecolor}{gray!20}
\definecolor{Gray}{gray}{0.8}

\newtheoremstyle{findingstyle}
  {0pt}   
  {0pt}   
  {\itshape}  
  {0pt}       
  {\bfseries} 
  {.}         
  {5pt plus 1pt minus 1pt} 
  {}          

\theoremstyle{findingstyle}
\newtheorem{findinner}{\textbf{Finding}}

\newenvironment{find}
  {\begin{shaded}\begin{findinner}}
  {\end{findinner}\end{shaded}}

\begin{document}

\acmYear{2026}\copyrightyear{2026}
\setcopyright{cc}
\setcctype[4.0]{by}
\acmConference[FSE Companion '26]{34th ACM Joint European Software Engineering Conference and Symposium on the Foundations of Software Engineering}{July 5--9, 2026}{Montreal, QC, Canada}
\acmBooktitle{34th ACM Joint European Software Engineering Conference and Symposium on the Foundations of Software Engineering (FSE Companion '26), July 5--9, 2026, Montreal, QC, Canada}
\acmDOI{10.1145/3803437.3805557}
\acmISBN{979-8-4007-2636-1/26/07}

\title{Towards Reliable Testing of Machine Unlearning}

\author{Anna Mazhar}
\affiliation{%
	\institution{Cornell University}
	\city{Ithaca}
	\country{NY, USA}
}

\author{Sainyam Galhotra}
\affiliation{%
	\institution{Cornell University}
	\city{Ithaca}
	\country{NY, USA}
}

\begin{abstract}
Machine learning components are now central to AI-infused software systems, 
from recommendations and code assistants to clinical decision support. 
As regulations and governance frameworks increasingly require deleting sensitive data 
from deployed models, \emph{machine unlearning} is emerging as a practical 
alternative to full retraining. However, unlearning introduces a 
software quality-assurance challenge: under realistic 
deployment constraints and imperfect oracles, 
how can we test that a model no longer relies on targeted information? 
This paper frames \emph{unlearning testing} as a first-class 
software engineering problem. We argue that practical unlearning tests 
must provide (i) thorough coverage over proxy and mediated influence pathways, 
(ii) debuggable diagnostics that localize where leakage persists, 
(iii) cost-effective regression-style execution under query budgets, and 
(iv) black-box applicability for API-deployed models. 
We outline a causal, pathway-centric perspective, \emph{causal fuzzing}, 
that generates budgeted interventions to estimate residual direct and 
indirect effects and produce actionable ``leakage reports''. 
Proof-of-concept results illustrate that standard attribution checks 
can miss residual influence due to proxy pathways, 
cancellation effects, and subgroup masking, motivating 
causal testing as a promising direction for unlearning testing.
\end{abstract}

\maketitle

\vspace{-2mm}
\section{Introduction}
\label{sec:intro}

Machine learning models increasingly power features in \emph{AI-infused software systems} including
user-facing and backend capabilities such as search, recommendations, code assistants, and clinical triage.
Engineering these systems differs from traditional software because model behavior depends on data,
training pipelines, and deployment constraints that evolve over time \cite{AmershiICSE19}.
Testing and quality assurance for ML components has consequently become an active topic in software engineering research,
spanning test generation, adequacy criteria, and oracle challenges \cite{ZhangTSE22MLTesting,BarrTSE15Oracle}.

These systems are often trained on 
proprietary and user-generated data that often contains sensitive information. 
Regulations such as the GDPR and CCPA now empower users to request the
 removal of their data~\cite{GDPRArt17,CCPAOAG}, forcing organizations to ensure that models can 
 truly ``forget'' specified information. Beyond regulatory compliance,  
 robust unlearning is vital for mitigating bias that undermines trust 
 and safety, and it plays a key role in post-deployment debugging by 
 exposing and correcting spurious correlations.

Retraining models from scratch after every deletion request is often impractical for large-scale systems.
\cite{bourtoule2021machine}
This motivates the community to study approximate unlearning methods that update model parameters to
(approximately) erase targeted information.

While these methods can be more efficient than full retraining, they introduce a fundamental software
quality-assurance challenge: \textbf{under realistic deployment constraints, how can we test that unlearning has
eliminated all relevant traces of the forgotten data?} 
This question is hard for the same reason many testing problems are hard: the desired property
(``the model has forgotten $Z$'') rarely has a perfect oracle, so practitioners must rely on systematic,
indirect evidence \cite{BarrTSE15Oracle}.

Existing unlearning checks largely rely on data-sample removal or feature-level attribution.
 While useful, these approaches focus on surface associations and often fail to detect 
 residual influence that persists indirectly through correlated proxies or mediated pathways,
 precisely the failure modes that matter in post-deployment testing.

\vspace{-2mm}
\example{
    \label{ex:lung-disease}
    Consider a lung disease diagnosis model trained on patient data 
including X-rays, patient metadata, and clinical notes.
Suppose the model learns to associate \texttt{Hospital A} 
with higher disease risk
because that hospital tends to treat more severe cases. 
A regulator subsequently mandates unlearning of hospital identifiers 
to prevent institutional bias.
After unlearning, the model no longer reads visible hospital IDs or metadata.

\noindent\textit{Leakage through structured data.} 
Even after removing hospital identifiers, the model may still infer 
institutional affiliation from correlated demographic attributes. 
For instance, features such as residential zip-code or employment history 
often indirectly encode where a patient is likely to receive care, 
As a result, the model continues to exhibit biased predictions, 
even though direct identifiers have been stripped away.  

\noindent\textit{Leakage through unstructured data.} 
A similar issue may arise with imaging data. 
Hospitals may configure their scanners differently, for example, 
using distinct grid settings, contrast levels, or reconstruction parameters. 
Even after unlearning hospital identifiers, the model may still recognize 
these scanner-specific signatures and associate them with higher disease risk.  
}

Prior work on feature attribution and leakage analysis, 
including Shapley values and influence 
functions~\cite{lundberg2017unified,koh2017understanding,garima2020estimating}, 
can expose direct associations, but it does not verify whether 
residual influence persists through mediated or proxy pathways.
These findings suggest that indirect signals, 
both structured or unstructured, can still encode residual influence,
undermining the goal of unlearning.

\begin{figure*}[h!]
\centering
\includegraphics[width=\linewidth]{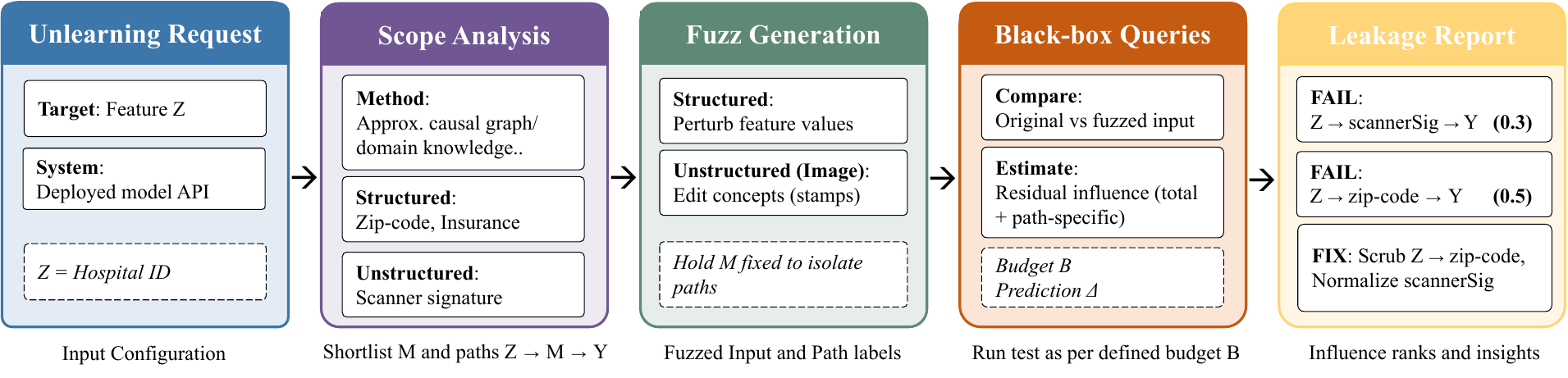}
\caption{Causal fuzzing workflow: from unlearning request to actionable leakage report via path-guided input generation}
\label{fig:intro-exp}
\vspace{-3mm}
\end{figure*}

This paper envisions a causal unlearning testing framework that moves beyond binary pass/fail judgments 
to form a feedback loop for improving unlearning methods. To be viable within routine 
software quality-assurance pipelines, such testing must satisfy four tightly coupled properties. 
\textbf{Thoroughness} requires detecting all practically relevant residual influence of the targeted 
information, including indirect effects that persist through correlated proxies or mediated 
causal pathways. \textbf{Debuggability} ensures that when residual influence is detected, 
tests explain where and how unlearning failed to help localize the features or causal 
pathways responsible for leakage to enable targeted remediation. \textbf{Cost-effectiveness} 
makes this process sustainable in practice, allowing tests to run repeatedly 
(e.g., per deletion batch or model release) by prioritizing high-yield tests and reusing 
artifacts without incurring prohibitive computational or human cost. Finally, \textbf{black-box 
applicability} ensures deployability in real systems by relying only on input–output 
behavior, even when model internals are inaccessible, as in common API-based settings.
These four properties follow directly from unlearning constraints in deployed ML systems.

To achieve these objectives, we propose framing unlearning testing as a \emph{causal influence} problem: 
the claim ``\(Z\) is forgotten'' should mean that both its \emph{direct} and \emph{indirect} effects on predictions are negligible. 
Specifically, we propose a ``causal fuzzing'' mechanism to systematically test models
and produce actionable,
pathway-centric diagnostic insights under black-box access (Figure~\ref{fig:intro-exp}). 

Our preliminary evidence on structured data shows that path-specific testing detects 
residual leakage that other methods miss, while remaining black-box. 
Looking ahead, we plan to extend the notion of mediator paths to natural
language  and image data using modality-specific perturbations.
More broadly, we highlight a range of open research questions 
(\S\ref{sec:future}) that call for community effort to advance rigorous,
efficient, and generalizable unlearning testing.

\vspace{-2mm}
\section{Background and Related Work}
\label{sec:related}

SE has studied systematic testing of ML components using metamorphic relations and test generation,
including DeepXplore~\cite{pei2017deepxplore}, DeepTest~\cite{tian2018deeptest}, DeepHunter~\cite{xie2018deephunter}.
Recent work motivates black-box testing under deployment constraints, e.g., via test-case diversity~\cite{aghababaeyan2021blackboxdiversity},
and advances metamorphic and contrastive testing for modern ML services~\cite{chen2024missmile,jiang2024costello,mu2025improving,wang2023distxplore}.
These techniques motivate our emphasis on efficiency and black-box applicability, but they do not target \emph{forgetting}.

\vspace{-1mm}
\smallskip\noindent\textbf{Unlearning Evaluation and Verification.}
Machine unlearning aims to update a model as if certain training data had never been used~\cite{bourtoule2021machine,izzo2021approximatedatadeletionmachine}.
A growing line of work studies how to evaluate/verify unlearning, including memorization-based metrics~\cite{Jagielski2023MeasuringFO},
stronger adversarial protocols~\cite{sommer2022athena,goel2023towards}, and evidence that verification can be fragile or introduce new risks
(e.g., dishonest providers and reconstruction attacks)~\cite{bertran2024reconstructionunlearning}.
Benchmarks and surveys further systematize evaluation (e.g., MUSE)~\cite{shi2024muse,ZhangTSE22MLTesting}.
However, these evaluations typically provide aggregate evidence and lack pathway-level, debuggable tests under black-box constraints.

\vspace{-1mm}
\smallskip\noindent\textbf{Influence and Leakage Analyses.}
A large body of work estimates data/feature influence via Shapley values, influence functions, and gradient/trajectory analyses~\cite{lundberg2017unified,koh2017understanding,garima2020estimating},
and uses attribution methods such as saliency maps and integrated gradients for unstructured models~\cite{simonyan2013saliency,sundararajan2017axiomatic}.
Complementary privacy and extraction work shows that models can retain and reveal training information~\cite{shokri2017membership,salem2018ml,salem2020updates,chen2021when,carlini2021extracting,carlini2023extracting}.
Overall, these approaches often quantify surface exposure or direct association and may miss mediated leakage through proxies.

\vspace{-3mm}
\section{Desired Properties of Unlearning Tests}
\label{sec:unlearning}

Machine unlearning introduces a \emph{post-deployment} quality-assurance problem for AI-infused software systems:
after an unlearning request, practitioners need \emph{testable} evidence that the model no longer relies on the
targeted information.
In deployed settings this evidence must be collected under limited observability (e.g., API-only access) and
imperfect oracles~\cite{AmershiICSE19,BarrTSE15Oracle}.
From a testing perspective, unlearning checks are useful only if they (i) achieve \emph{thoroughness} (high coverage)
to avoid false reassurance when influence persists through proxies, (ii) are \emph{debuggable} so
failures localize \emph{which} routes remain influential and guide remediation, (iii) are \emph{cost-effective} enough
to run routinely (e.g., per deletion batch or release), and (iv) remain \emph{black-box applicable} because many models
are accessible only via service interfaces.
These properties represent a deployment-driven minimal set of necessary
 conditions for unlearning tests to function as first-class software tests in deployed settings.

\vspace{-2mm}
\subsection{Thoroughness (Coverage)}
Unlearning tests should detect \emph{all practically relevant} residual influences.
For ML components, relying on a single simplistic coverage target 
is risky. Prior work shows that common DNN coverage criteria may correlate 
weakly with fault detection or model quality~\cite{YanFSE20Coverage,HarelCanadaFSE20NeuronCoverage}.
Analogously, an unlearning test suite that checks only whether the
\emph{target feature} remains explicitly accessible can miss proxy-mediated influence.

Consider the testing implications for our hospital unlearning scenario 
(exp~\ref{ex:lung-disease}). A surface-level test might verify that 
hospital identifiers are no longer directly readable from model inputs or outputs. 
However, this is incomplete: the model may still route hospital-specific bias through
demographic proxies (service-area indicators) or subtle imaging artifacts.

To ensure high coverage, testing must therefore probe these indirect channels systematically. 
For demographic features (structured), this involves testing whether zip codes, insurance types, 
or referral patterns still enable hospital inference. For imaging data (unstructured), verification 
requires checking if scanner-specific noise patterns continue to influence predictions in hospital-correlated ways.
In practice, this is also an \emph{oracle} challenge: we rarely have 
a perfect ``forgotten/not-forgotten'' oracle in this domain, motivating tests that 
leverage structured invariants (e.g., metamorphic-style relations) 
rather than relying on labels alone~\cite{BarrTSE15Oracle,ChenMT98}.

Effective unlearning verification therefore cannot rely on feature-by-feature checks.
It requires a \emph{pathway-centric} approach that traces how target information flows
through the model and detects both direct dependencies and complex mediated effects.

\vspace{-2mm}
\subsection{Debuggability}
Unlearning testing must detect both residual influence and guide mitigation.
Debuggability requires actionable localization of \emph{where} leakage persists, analogous to
fault localization and delta debugging
\cite{JonesICSE02Tarantula,ZellerTSE02Delta,ZellerFSE02CauseEffect}.
In our hospital example, removing explicit identifiers may not 
eliminate institution-specific bias.
A non-debuggable test yields only a binary verdict 
(e.g., ``hospital bias detected''), leaving developers uncertain whether 
leakage arises from demographics, imaging artifacts, or other proxies.

A debuggable test should report \emph{which pathways} 
remain influential (e.g., leakage through scanner-specific grid patterns 
in image corners, with secondary leakage through geographic proxies).
Such localization enables targeted interventions, such as preprocessing 
to normalize scanner signatures, or 
applying unlearning focused on specific mediators.

\vspace{-2mm}
\subsection{Cost-effectiveness}
For unlearning verification to be practical, 
particularly in large-scale or frequently 
updated models, it must be cost-effective. 
Comprehensive tests may require thousands of probes spanning different modalities, 
languages, and inference patterns, which can be prohibitively expensive 
even for verifying a single datapoint. 
In large language models, the combinatorial explosion of
possible prompts and contexts can compound into substantial costs.
Hence, balancing cost-effectiveness with thoroughness is crucial.
This motivates prioritization mechanisms (which pathways to test first), 
reuse of test artifacts across versions, and guided exploration strategies akin to 
fuzzing's budgeted search for high-value tests~\cite{LemieuxCCS18FairFuzz}.

\vspace{-2mm}
\subsection{Black-Box Applicability}
AI-infused systems are often API-only, with limited internal access.
Consequently, unlearning tests should support black-box setting using only inputs and outputs.
This aligns with a key software testing principle: 
feasibility under production observability~\cite{AmershiICSE19}.

Black-box applicability also supports governance and audit contexts 
where independent validation may be required, and aligns with operational 
monitoring expectations in risk-management frameworks~\cite{NISTAIRMF,ISO23894,EUAIActArt11,EUAIActArt72}.
At the same time, black-box constraints create explicit tradeoffs: 
they can reduce diagnosability and thoroughness unless the testing strategy 
leverages structure to compensate.

\vspace{-2mm}
\section{Causal Perspective on Unlearning}
\label{sec:causal-testing}

Machine unlearning introduces a post-deployment testing problem: after an unlearning request,
practitioners need evidence that the model no longer relies on the targeted information $Z$,
despite limited observability (often API-only).
We propose a \emph{causal testing} view: unlearning is successful only if the \emph{direct and indirect}
causal influence of $Z$ on the model output $Y$ is negligible.

\vspace{-1mm}
\smallskip\noindent\textbf{Test Oracle.}
We represent the deployed predictor as part of a structural causal model (SCM), where nodes correspond to
observed inputs/features, intermediate representations, and outputs.
Unlearning verification reduces to testing whether interventions on $Z$ can still change $Y$, either
directly or through mediators.
This yields a concrete oracle: if $Z$ is forgotten, then intervening on $Z$ and propagating its downstream
causal consequences should not meaningfully change $Y$.
For each intervention, model outputs on original and intervened inputs are compared via the absolute
change in prediction score (or class probability), and expected residual influence is estimated by
Monte Carlo averaging.
Residual influence is thresholded to decide whether unlearning is complete.
The oracle assumes access to causal structure as a directed acyclic graph and the ability to sample
realistic intervention values, while functional dependencies remain unknown.

\smallskip\noindent\textbf{Causal fuzzing workflow (how it operates).}
Figure~\ref{fig:intro-exp} summarizes the causal fuzzing workflow.
Given a target $Z$, we first generate a \emph{budgeted} 
test suite of interventions that probes candidate influence routes from $Z$ to $Y$:
(i) \emph{Scope \& candidates:} identify a small set of likely mediators/proxies $M$ (from the causal graph,
domain knowledge, or simple proxy screening);
(ii) \emph{Interventions:} generate fuzz inputs (feature perturbations for structured data; concept edits for images/text) 
by intervening on $Z$ (and optionally blocking or
fixing selected mediators $M$) to isolate total and path-specific effects.
(iii) \emph{Black-box queries:} query the model on the original vs. intervened inputs and estimate residual
effects (total and/or path-specific) under a fixed query budget;
(iv) \emph{Leakage report:} rank failing tests by effect size and report \emph{which mediator/path} explains
the residual influence. 
The report should be structured as ranked entries e.g., 
$Z\!\rightarrow\!\texttt{BMI}\!\rightarrow\!Y$ with a
high estimated effect and \texttt{BMI} flagged for inspection.

Approximate structure may come from domain knowledge, or causal
discovery (e.g., LiNGAM~\cite{shimizu2006linear}).
In practice, coarse structure can target likely mediator routes and support
refutation/robustness checks to detect some graph misspecifications~\cite{sharma2020dowhyendtoendlibrarycausal}.
We discuss reporting graded assurance under graph uncertainty in Section~\ref{sec:future}.

\smallskip\noindent\textbf{How this achieves the desired properties.}
Causal fuzzing satisfes the desired properties (\S\ref{sec:unlearning}) as follows.
\emph{High coverage} comes from targeting \emph{paths} from $Z$ to $Y$ (direct and mediated) rather than
enumerating features in isolation.
\emph{Debuggability} comes from attributing a failure to specific mediator sets or paths (e.g.,
$Z\!\rightarrow\!M\!\rightarrow\!Y$), producing actionable localization.
\emph{Cost-effectiveness} comes from budgeted testing: prioritizing a small number of high-risk mediators
(e.g., by causal proximity, proxy strength, domain risk, or prior failures) 
instead of exhaustively perturbing all features
and reusing test artifacts across model versions, akin to guided fuzzing 
under a fixed budget~\cite{LemieuxCCS18FairFuzz}.
\emph{Black-box applicability} holds because causal testing requires only model 
queries on interventions.

\vspace{-2mm}
\subsection{Proof-of-Concept: Structured Data}
\label{sec:poc-structured}

\smallskip\noindent\textbf{Goal and setup.}
To validate feasibility, we ran proof-of-concept experiments on predictors 
representative of decision and health informatics software,
trained on structured datasets commonly used in SE fairness-testing literature~\cite{chen2024fairness}: 
\textit{Adult Income} and \textit{Drug Consumption} (real-world)~\cite{adult_2,drug_consumption},
and \textit{Heart-Disease} (semi-synthetic).
We compared against baseline testing approaches:
\texttt{permutation importance}~\cite{fisher2019modelswrongusefullearning},
\texttt{SHAP}~\cite{lundberg2017unified}, 
and fairness metrics~\cite{bellamy2018aifairness360extensible}.
We chose targets $Z$ that are (i) explicitly sensitive (\texttt{gender}, \texttt{age}) or
(ii) plausibly causal for outcomes (\texttt{smoking}, \texttt{BMI}).

    For Adult Income, we used a coarse causal graph specified 
    based on domain knowledge~\cite{silvia2019pathspecific}; 
    for Drug Consumption, we initialized a graph with LiNGAM~\cite{shimizu2006linear}
    and retained only domain-plausible edges for testing; 
    for the semi-synthetic Heart-Disease dataset, we used the known SCM structure.
    We treat these graphs as coarse structural hypotheses used to target tests,
    not as guaranteed ground-truth causal structures.
    Our goal is not exact causal identification, but evaluating whether testing
    remains useful without perfect causal knowledge.
    We discuss reporting graded assurance under graph uncertainty in Section~\ref{sec:future}.

\smallskip\noindent\textbf{Unlearning baseline (to generate post-unlearning models).}
We use \emph{feature removal} as a baseline unlearning operation (retrain without $Z$).
This mirrors a common mitigation in practice and provides a clean setting to evaluate testing effectiveness:
$Z$ is no longer directly available, yet its influence may persist through correlated proxies or mediated routes.
This simplified proxy for unlearning enables controlled evaluation of residual indirect effects, 
but it does not capture more realistic unlearning methods, which we leave for future work.
Our focus is the \emph{testing} strategy (causal fuzzing), rather than proposing a new unlearning algorithm.

\vspace{-1mm}
\subsubsection{Proxy pathways.}
A frequently observed failure mode is that unlearned information persists through alternative causal routes.
After unlearning \texttt{smoking} in \textit{Heart-Disease}, \texttt{SHAP} and \texttt{permutation importance}
suggested that \texttt{smoking} influence was negligible.
However, causal fuzzing revealed substantial residual influence
mediated through \texttt{blood pressure} and \texttt{BMI}.
This illustrates that removing $Z$ does not prevent 
the model from re-expressing its influence via proxies
even when surface-level attribution looks negligible.

\vspace{-2mm}
\subsubsection{Cancellation illusion.}
We observed that aggregate influence measures can underestimate residual 
dependence when opposing mediated effects cancel.
In \textit{Drug Consumption}, \texttt{Education} appeared weak overall in baseline evaluations.
Path-specific analysis using causal fuzzing revealed why: two strong 
but opposing mediated pathways.
Education simultaneously lowered risk by increasing health awareness 
and conscientiousness (\textit{Cscore}),
while increasing risk through elevated social exposure and extraversion (\textit{Escore}).
Although these effects nearly cancel in aggregate, 
the model retains both dependencies; small distributional
shifts can disrupt the balance and re-activate residual influence.
This blind spot motivates path-specific tests as a more faithful oracle for unlearning.

\vspace{-2mm}
\subsubsection{Subgroup masking.}
\begin{figure}[ht]
\centering
\vspace{-1mm}
\begin{subfigure}[b]{0.52\linewidth}
    \centering
    \includegraphics[width=\columnwidth]{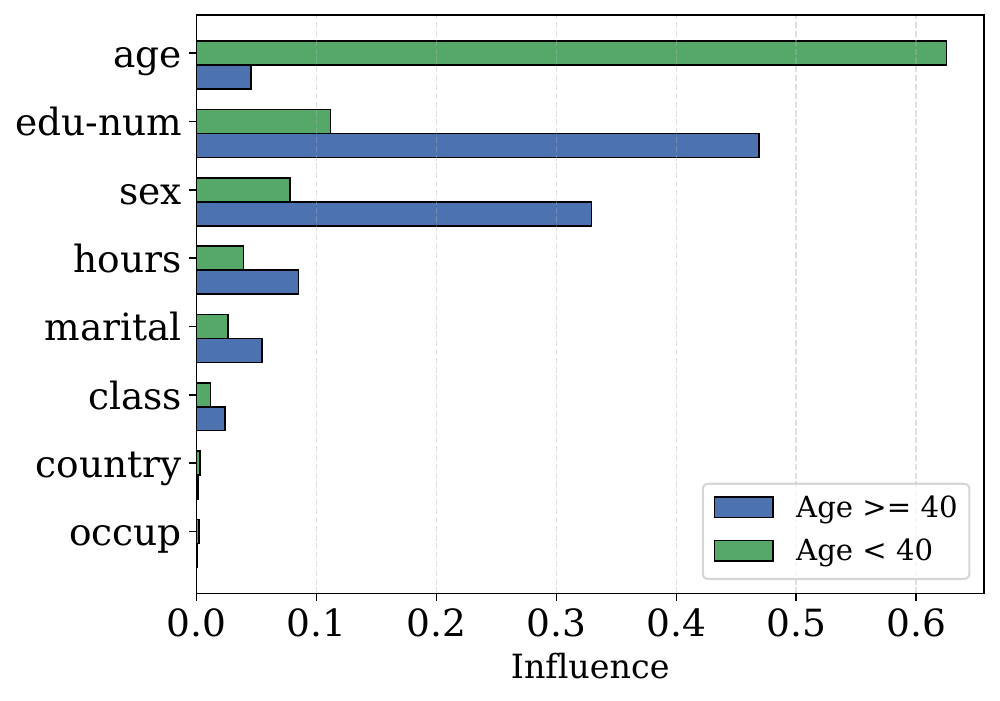}
    \caption{Indirect effect of \texttt{Age} vs all features in Adult Income.}
    \label{fig:adult-breakdown}
\end{subfigure}
\quad
\begin{subfigure}[b]{0.43\linewidth}
    \centering
    \includegraphics[width=\columnwidth]{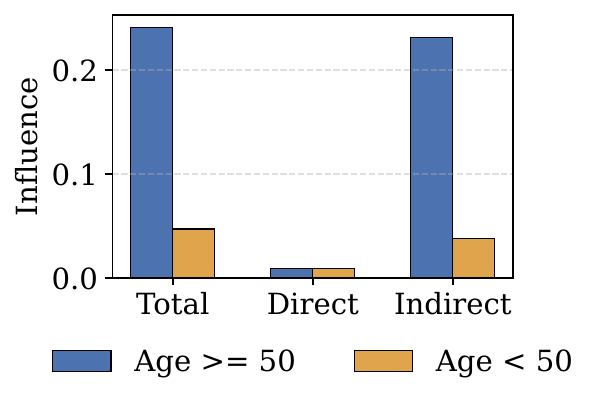}
    \caption{Indirect effect of \texttt{BMI} on Heart Disease in older group.}
    \label{fig:heart-breakdown}
\end{subfigure}
\vspace{-6mm}
\caption{Mediated effects by age group.}
\vspace{-6mm}
\label{fig:subfigures}
\end{figure}

Global testing can obscure subgroup-specific vulnerabilities.
In \textit{Adult Income}, \texttt{Age} appeared only moderately 
influential overall, yet among individuals under 40
it became a dominant factor (Figure~\ref{fig:adult-breakdown}).
Likewise, in \textit{Heart-Disease} after unlearning \texttt{BMI}, indirect influence was substantially stronger for
older individuals and mediated through \texttt{blood pressure} (Figure~\ref{fig:heart-breakdown}).
In both cases, \texttt{SHAP} and \texttt{permutation importance} 
failed to capture this subgroup-specific influence variation.

\smallskip\noindent\textbf{Summary.}
These proof-of-concept results highlight that conventional verification can miss indirect, cancelling, and
subgroup-specific effects, creating blind spots in unlearning assessment.
By contrast, causal fuzzing yields pathway-level evidence that directly supports \emph{coverage} and
\emph{debuggability}, while remaining compatible with black-box model access.
As with other testing techniques, passing all unlearning tests does not prove absence of leakage, 
but systematically reduces known high-risk failure modes.
Future work will scale the approach to higher-dimensional domains and study cost/coverage tradeoffs under
explicit query budgets.

\section{Open Challenges and Future Directions}
\label{sec:future}

Our proof-of-concept illustrates that causal, pathway-centric tests can expose verification blind spots,
but several challenges remain before such tests become routine regression artifacts in SE pipelines.

\smallskip\noindent\textbf{Imperfect causal knowledge.}
An open question is formalizing when such tests remain informative 
under graph uncertainty, i.e., how coverage and
localization degrade as causal knowledge becomes partial.
Promising directions include uncertainty-aware testing (e.g., testing families of plausible graphs) and
reporting assurance levels rather than binary verdicts.

\smallskip\noindent\textbf{Budgeted regression testing.}
To support repeated unlearning in practice, future work should develop regression-style test suites that
reuse perturbation artifacts across releases, prioritize high-risk pathways, and make cost/coverage tradeoffs
explicit under budgets.

\smallskip\noindent\textbf{Unstructured and foundation models.}
Extending causal fuzzing beyond tabular data requires reliable modality-specific interventions (e.g., concept-level
perturbations for images/text) and validity checks to ensure edits are realistic.
In text setting, a key open problem is defining text-level mediators that remain meaningful under paraphrase and other
surface changes. Another is designing realistic interventions that preserve task intent while perturbing
candidate mediators, and deciding how residual influence should be estimated under a fixed prompt/query budget.
For LLMs/VLMs, an additional challenge is prompt sensitivity; standardized prompt suites and pathway-oriented
tests are needed to provide comparable evidence under black-box access.
\section{Conclusion}
\label{sec:conclusion}

In this paper, we have framed machine unlearning testing as a 
critical concern for for AI-infused software systems.
We have identified key properties that practical unlearning tests must possess, 
including thoroughness, debuggability, efficiency, and black-box compatibility.
Through our exploration of causal testing methods, 
we have demonstrated their potential to uncover 
residual influences and provide actionable insights for practitioners. 
Our vision is to establish rigorous testing 
frameworks that ensure genuine data removal while 
addressing the complexities of modern ML models.

\bibliographystyle{ACM-Reference-Format}
\bibliography{ref}
\end{document}